\documentclass[preprint,showpacs,preprintnumbers,amsmath,amssymb]{revtex4}

\usepackage{graphicx}
\usepackage{dcolumn}
\usepackage{bm}
\usepackage[]{amssymb,epsfig,color}
\newcommand{\be}{\begin{equation}}
\newcommand{\ee}{\end{equation}}
\begin{document}
\title{Chaotic dynamics and spin correlation functions in a chain of nanomagnets.}

\author{L. Chotorlishvili$^{1,3}$,    Z. Toklikishvili$^{2}$,   A. Komnik$^{3}$, J. Berakdar$^{1}$ }

\affiliation{1 Institut f\"ur Physik, Martin-Luther Universit\"at
Halle-Wittenberg, Heinrich-Damerow-Str.4,
D-06120 Halle, Germany\\
 2 Physics Department of the Tbilisi State University,
                Chavchavadze av.3, 0128, Tbilisi, Georgia\\
 3 Institut f\"ur Theoretische Physik Universit\"at Heidelberg,
Philosophenweg 19, D-69120 Heidelberg}
\begin{abstract}
We study a chain of coupled nanomagnets in a classical approximation. We show
 that the infinitely long chain of coupled nanomagnets  can be  equivalently mapped
  onto an effective one-dimensional  Hamiltonian  with a fictitious time-dependent perturbation.
  We establish a connection between the dynamical characteristics of the classical system and spin correlation time. The decay rate for the spin correlation functions turns out to  depend logarithmically on the maximal Lyapunov exponent. Furthermore, we discuss the non-trivial role of the  exchange anisotropy within the chain.

\end{abstract}
\pacs{05.45.Gg, 75.50.Xx, 75.78.Jp, 05.10.Gg}
\maketitle
\section{Introduction}

Nanoscale magnetic structures have promising applications  as basic elements in
 future nanoelectronics devices and are frequently discussed
 in the context of  quantum
information processing. The principal challenge of the
quantum information technology is finding an efficient procedure
for the generation and manipulation of the many-qubit entangled
states. Those can be realized on the basis of e.~g. Rydberg atoms
located in  optical quantum cavities
\cite{Raimond,Amico,Mabuchi}, Josephson junctions\cite{You}, or
ion traps \cite{Blatt}. One very promising realization is based on
 single molecular nanomagnets
(SMMs)\cite{Sessoli,Gunther,Friedman}. These are molecular
structures with a large effective spin. A prototypical
representative of this family of compounds is, for example,
$Mn_{12}$ acetate in which $S=16$.  Molecular nanomagnets show a number of
interesting phenomena that have been in the focus of theoretical
and experimental research during the last two
decades.\cite{Sessoli,Gunther,Friedman,Hernandez,Lionti,Chudnovsky,Thomas,Wernsdorfer,Hennion,Nature,
Tiron,Ardavan} For instance, SMMs show a bistable behavior as a
result of the strong uniaxial anisotropy \cite{Sessoli} as well as
a tunneling of the magnetization.\cite{Gunther} An attractive feature
for information storage is the
large relaxation time of molecular nanomagnets.\cite{Chudnovsky}

SMMs are usually modelled by spin chain Hamiltonians augmented by different kinds of interaction terms responsible for different compounds. These interaction contributions are highly non-trivial and in most cases are anisotropic. This makes the analytical treatment very cumbersome  calling for
 efficient theoretical approaches. In this paper we  investigate the properties of a classical spin chain coupled by anisotropic
exchange interactions. This case is relevant not only for  chains of exchange coupled SMMs,\cite{Christou} but also for several other realistic
physical problems. Those include weakly coupled antiferromagnetic rings
\cite{Candini} and large spin multiples coupled by Dzyaloshinskii-Moriya (DM) exchange interaction\cite{Stamatatos}.  We will demonstrate that even for
multidimensional complex physical systems it is still possible to obtain
analytical results using special mathematical technique presented in \cite{Berakdar}.
Its applicability to the SMMs is shown in \cite{Chotorlishvili,Schwab}.
In the first step one evaluates the Lyapunov exponents and the spin correlation functions
for the system. Having done that one can then extract information on the properties of the system beyond the classical limit. For example, there is a deep connection between the classical Lyapunov exponents and the quantum Loschmidt echo \cite{Iomin}, which is a natural measure of the
quantum stability and of the fidelity of quantum teleportation
\cite{Bruss}. If the Lyapunov regime is reached for a quantum system,
then the decay rate for the teleportation fidelity can be identified via the
Lyapunov exponent. Formal criteria for the Lyapunov regime
\cite{Jacquod} in case of a chain of SMM can be estimated from the relation $\lambda< J^2/\Delta$, where $\lambda$ is the Lyapunov exponent,  $\Delta$ is the mean level spacing
and $J$ is the exchange interaction constant between SMM spins (which
prohibits the integrability of the system leading thus to the
chaotic dynamics). Furthermore, it can be shown that the spin correlation
functions can be expressed through the Lyapunov exponent.

The paper is organized as follows. In Section II we give a brief exposition of our model and present the details of the  principal investigation technique. After that in Section III we discuss the relation between the relevant spin correlation functions and their classical analogs. Section IV contains the treatment of the system in the chaotic domain. Finally, the conclusions section summarizes our findings. All necessary mathematical details are presented in Appendices.

\begin{widetext}
\section{Theoretical formulation}
The prototype model Hamiltonian for the exchange coupled SMM
is \be \label{eq:hamiltonian}
H=J\sum\limits_{n}S_{n}^{z}S_{n+1}^{z}+g\sum\limits_{n}\bigg(S_{n}^{x}S_{n+1}^{x}
+S_{n}^{y}S_{n+1}^{y}\bigg)+\beta\sum\limits_{n}\bigg(S_{n}^{z}\bigg)^{2}
\, , \ee where $J$ and $g$ are exchange interaction constants,
$\beta =-DS^{2}$ stands for the anisotropy barrier height of the
system. For the prototypical $Mn_{12}$ acetates \cite{Sessoli}
$D\sim 0.7K$ sets the value of the barrier parameter \cite{Schwab}
and $S^{x,y,z}$ are spin projection operators of SMM. Due to the large
spin of the SMM  (see Introduction) analytical quantum
mechanical treatment of the model (\ref{eq:hamiltonian}) is
hardly accessible. To make progress we choose the semi-classical
parametrization as follows: \be\label{eq:parametrization}
S_{n}^{z}=\cos\theta_{n}, S_{n}^{x}=\sin\theta_{n}\cos\varphi_{n},
S_{n}^{y}=\sin\theta_{n}\sin\varphi_{n} .
 \ee
Then the Hamiltonian (\ref{eq:hamiltonian}) can be rewritten in the more convenient
 form
 \be\label{eq:ham}
 H=J\sum\limits_{n}\cos\theta_{n}\cos\theta_{n+1}+g\sum\limits_{n}\sin\theta_{n}\sin\theta_{n+1}\cos(\varphi_{n+1}-\varphi_{n})
 +\beta\sum\limits_{n}\cos^{2}\theta_{n}.
 \ee
 Our aim is the evaluation of the correlation functions and the study
 of the spin dynamics governed by the Hamiltonian (\ref{eq:ham}). Since this is highly nonlinear problem, it cannot be done in a simple and direct way. However, one can rigorously show, that there is a
 direct map between the chain of SMM and  a one-dimensional ($1D$) model Hamiltonian  with a fictitious time-dependent external perturbation.

 The equilibrium state for the model (3) satisfies the minimum condition of the infinite-dimensional functional
$H[\theta,\varphi]$:
\begin{eqnarray}\label{eq:minima}
&&\frac{\partial H}{\partial \theta_{n}}=0,~~~~\frac{\partial
H}{\partial \varphi_{n}}=0, \\
&&n=1,2\ldots\infty. \nonumber
\end{eqnarray}
Considering the Hamiltonian (\ref{eq:ham}) we retain only the
first order terms of the anisotropy parameter
$\varepsilon=(J-g)/2g \ll 1$. Then, after straightforward but
laborious calculations we deduce from (\ref{eq:minima})  that the following relations hold
\begin{eqnarray}\label{eq:atab}
&&S_{n+1}=(-1)^{m}\big[S_{n}-\beta\sin(2\theta_{n})(1-\varepsilon\cos(2\theta_{n}))\big],\\
&&\theta_{n+1}=(-1)^{m}\theta_{n}+\pi \nu +(-1)^{\nu}\arcsin
S_{n+1},\\
&& \varphi_{n+1}=\varphi_{n}+\pi m,~~~m=0,1~~~
\nu=0,1,  \nonumber
\end{eqnarray}
where $S_{n+1}=\sin(\theta_{n+1}-\theta_{n})$. The index $\nu$
refers to the two possible solutions when inverting the trigonometric
expression for $S_{n+1}$. Depending on the sign of the re-scaled
barrier height $\beta\rightarrow \mp\ \beta /g$ the
index $m=0,1,$ defines the energy minimum condition. For convenience
we will use positively defined $\beta
>0$ and consequently $m=0$ in our calculations.  The above result
is just a recurrence relation in the form of the explicit
map
$\big(S_{n+1},\theta_{n+1}\big)=\hat{T}\big(S_{n},\theta_{n}\big)$.
    Our idea is to find a Hamiltonian model which is equivalent to (\ref{eq:atab}). Let us consider the following perturbed $1D$  Hamiltonian
system:
\begin{eqnarray}\label{eq:hamsystem}
&&H=H_{0}(s)+\beta V(\theta)T
\sum\limits_{n=-\infty}^{+\infty}\delta(t-nT) \, , \nonumber \\
&&H_{0}(S)=\nu\pi S+(-1)^{\nu}\bigg(S\arcsin S+\sqrt{1-S^2}\bigg)\, ,\\
&& V(\theta)=-\bigg(\cos^{2}-\frac{\varepsilon}{4}\cos^{2}2\theta
\bigg). \nonumber
\end{eqnarray}
The respective Hamiltonian equations read
\begin{eqnarray}\label{eq:motion}
&&\frac{dS}{dt}=-\frac{\partial H}{\partial \theta}=-\beta
V'(\theta)T
\sum\limits_{n=-\infty}^{+\infty}\delta(t-n T),\nonumber\\
&& \frac{d \theta}{d t}=\frac{\partial H}{\partial S}=\omega (S),
\\
&& \omega_{\nu}(S)=\pi \nu+(-1)^{\nu}\arcsin s \, . \nonumber
\end{eqnarray}
Their integration simplifies due to the presence of the delta function in the
perturbation term because the evolution operator
$(\bar{S};\bar{\theta})=\hat{T}(S;\theta)$ splits in a pulse-induced
$\hat{T}_{\delta}$ and free evolution terms $\hat{T}_{R}$:
\begin{eqnarray}\label{eq:evoperator}
&&\hat{T}=\hat{T}_{R}\otimes\hat{T}_{\delta}, \nonumber\\
&&\bar{S}\equiv
S(t_{0}+t-0);~~~~\bar{\theta}\equiv\theta(t_{0}+T-0),\\
&&S\equiv S(t_{0}-0);~~~\theta\equiv\theta(t_{0}-0).\nonumber
\end{eqnarray}
For operator of free evolution we simply have
\be\label{eq:tr}
\hat{T}_{R}(S;\theta)=(S;\theta+\omega_{\nu}(S)T). \ee The explicit
form of the pulse-induced evolution operator
$\hat{T}_{\delta}$ can be derived after an integration of the system
(6) on the small time interval $(t_{0}-0,t_{0}+0)$ around $t_0$ where the pulse is applied
\begin{eqnarray}\label{eq:integration}
&&
S(t_{0}+0)-S(t_{0}-0)=\int\limits_{t_{0}-0}^{t_{0}+0}\dot{S}dt=-\int\limits_{t_{0}-0}^{t_{0}+0}\beta
\frac{\partial V(\theta)}{\partial \theta}T
\sum\limits_{k=-\infty}^{+\infty}\delta(t-k T)=-\beta T
\frac{\partial V(\theta)}{\partial \theta}, \nonumber \\
&&\theta(t_{0}+0)-\theta(t_{0}-0)=\int\limits_{t_{0}-0}^{{t_{0}+0}}\dot{\theta}dt=0.
\end{eqnarray}
Taking into account Eq.~(\ref{eq:integration}), for the pulse-induced
evolution operator we obtain
\be\label{eq:tdelta} \hat{T}_{\delta}=\bigg(S-\beta\frac{\partial
V(\theta)}{\partial \theta},\theta\bigg).
 \ee
 Combining  Eqs.~(\ref{eq:tr}) and (\ref{eq:tdelta}) the complete evolution picture can be expressed through the following map
 \begin{eqnarray}\label{eq:map}
 &&\big(\bar{S},\bar{\theta}\big)=\hat{T}\big(S,
 \theta\big)=\hat{T}_{R}\hat{T}_{\delta}(S,\theta)=\hat{T}_{R}\bigg(S-\beta
 T\frac{\partial V(S,\theta)}{\partial
 \theta},\theta\bigg)=\nonumber\\
 &&=\bigg(S-\beta\frac{\partial V(\theta)}{\partial
 \theta};\theta+\omega_{\nu}(\bar{S})\bigg) \, .
 \end{eqnarray}
Or in the explicit form:
\begin{eqnarray}\label{eq:explicit}
&&S_{n+1}=S_{n}-\beta\bigg(\sin2\theta_{n}-\frac{\varepsilon}{2}\sin
4\theta_{n}\bigg), \nonumber\\
&&\theta_{n+1}=\theta_{n}+\omega_{\nu}(S_{n+1}),\\
&&\omega_{\nu}(S_{n})=\pi \nu +(-1)^{\nu}\arcsin S_{n}, \nonumber
\end{eqnarray}
For the details of the derivation of Eq.(14) see \textbf{ Appendix
A}.

This result is obtained for the  kicked Hamiltonian model with
$T=1$ and it  matches exactly the recurrence relations in
Eqs.~(\ref{eq:atab}) obtained for the SMM chain. Such an analogy
is quite important since the infinite-dimensional nonlinear system
(\ref{eq:ham}) is now equivalent to the $1D$ Hamiltonian model. We
note, the discrete time in the perturbation term
(\ref{eq:hamsystem}) is fictitious and corresponds to the number
of the spins in the chain (\ref{eq:hamiltonian}).

\section{Spin dynamics and correlation functions}
We will proceed with the
equivalent $1D$ Hamiltonian model with fictitious time dependent
perturbation (\ref{eq:hamsystem}), which is more convenient than the
multidimensional nonlinear model (\ref{eq:ham}). Due to the
nonlinearity of the model (\ref{eq:hamsystem}) we should expect a
rich and a complex dynamics.
In particular, our purpose is to establish a connection between the chaotic
dynamics and the decay rates of the spin correlation functions.
As a first step
we construct the Jacobian matrix of the map
(\ref{eq:explicit}): \be\label{eq:M}\hat{M}=\left(\begin{array}{l}
\frac{\partial
\bar{S}}{\partial S}~~ \frac{\partial \bar{S}}{\partial \theta}\\
\frac{\partial \bar{\theta}}{\partial S}~~\frac{\partial
\bar{\theta}}{\partial\theta}\end{array}\right)=\left(\begin{array}{l}1~~~~~~~~-\beta
V''(\theta)\\\omega'(\bar{S})~~1-\beta\omega'(\bar{S})V''(\theta)\end{array}\right).\ee
The Lyapunov exponents can be evaluated as eigenvalues of this
matrix
and are thus given by \be\label{eq:liapynov}
\lambda_{1,2}=1+\frac{K}{2}\pm\sqrt{\big(1+\frac{1}{2}K\big)^{2}-1}
\, , \ee where \be
\label{eq:K}K=-\beta\omega'_{\nu}(\bar{S})V''(\theta) \,  \ee is
the chaos parameter \cite{Zaslavsky}. From
Eqs.~(\ref{eq:liapynov}) and (\ref{eq:K}) for the chaos parameter $K$
we find the following simple relation
\be\label{eq:K1}K=-\frac{2\beta(\cos2\theta-\varepsilon\cos4\theta)}{\sqrt{1-S{2}}}
\, . \ee The dynamics is expected to be chaotic if
$K>0,~~\lambda_{1}>1$ or $K<-4,~~\lambda_{2}<-1$. Therefore from
Eq.~(\ref{eq:K1}) we obtain the relevant intervals for the angle
variable
\begin{eqnarray}\label{eq:theta}
&&\theta\in\bigg(\pi
m+\frac{1}{2}\arccos\Big(\frac{1}{2\varepsilon}-\sqrt{\frac{1}{4\varepsilon^2}+2}\Big);(m+1)\pi-
\frac{1}{2}\arccos\Big(\frac{1}{2\varepsilon}-\sqrt{\frac{1}{4\varepsilon^2}+2}\Big)\bigg),\nonumber\\
&&m=0,\pm1\ldots .
\end{eqnarray}
In the isotropic case
 $J\approx g$, $\varepsilon\rightarrow 0$, this leads to
 \be\label{eq:theta1}
 \theta\in\bigg(\pi
 m+\frac{\pi}{4};(m+1)\pi-\frac{\pi}{4}\bigg),~~~~
 m=0,\pm1\ldots .
 \ee
 \begin{figure}[t]
 \centering
  \includegraphics[width=10cm]{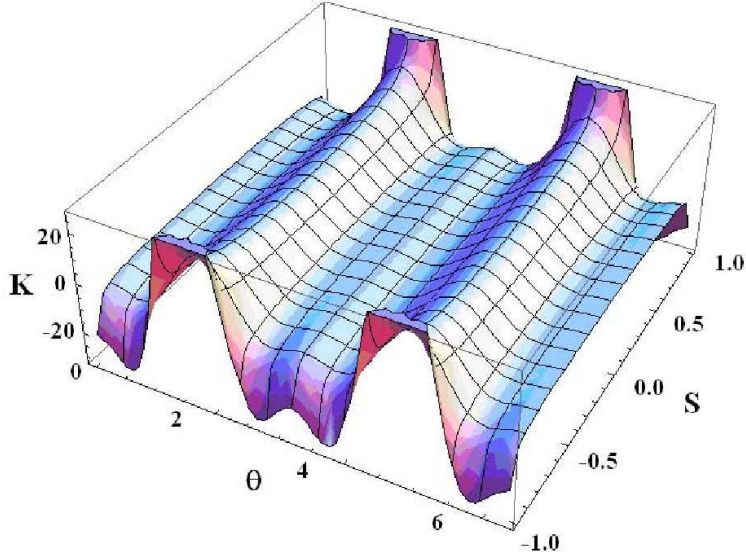}
  \caption{Color online. The parameter $K(\theta,S)$  signifying chaotic behavior
    plotted according to Eq. (\ref{eq:K1})  for the following values of the parameters $\beta=3$,$\varepsilon=0.5$.} \label{Fig:1}
\end{figure}
Equation (19) defiance the width of the chaotic domain where
parameter of chaos (18) is larger than one $K>1$. Obviously, the
width of the chaotic domain  depends on the values of anisotropy
and easy to see that in case of small anisotropy area of chaos is
narrower than in zero anisotropy case
$\Delta\theta(\varepsilon=0)>\Delta\theta(\varepsilon<<1)$.
Therefore we conclude  that small anisotropy leads to the less
chaotic regime.

From the parameter of the chaos (\ref{eq:K1}) (also plotted in
Fig.~\ref{Fig:1}) we conclude that the phase space of the system
consists of  domains corresponding to a regular and a  chaotic motion.
Later we will use the random phase approximation, which is valid precisely in the latter domain.\cite{Zaslavsky}

In order to obtain explicit expressions for the spin correlation
functions we rewrite the recurrence relations in Eqs.~(\ref{eq:explicit}) in
the following form \begin{eqnarray}
\label{eq:requrent}&&\theta_{n+1}=
\theta_{n}+\omega(s_{n+1})=\theta_{n}+\omega_{\nu}\big(S_{n}-\beta
V'(\theta)\big)=\nonumber\\
&&=\theta_{n}+\omega_{\nu}(S_{n})-\beta
V'(\theta)\omega'_{\nu}(S_{n})=\\
&&=\theta_{n}+\omega_{\nu}(S_{n})-\beta\bigg(\sin2\theta_{n}-
\frac{\varepsilon}{2}\sin4\theta_{n}\bigg) \, . \nonumber
\end{eqnarray}
For the angular variable we infer the self-consistent
recurrence relation  \be\label{eq:requrent1}
\theta_{n+1}=\theta_{n}+\omega_{\nu}(S_{n})-\beta\bigg(\sin2\theta_{n}-
\frac{\varepsilon}{2}\sin4\theta_{n}\bigg).\ee The correlation
function is given by
\begin{eqnarray}\label{eq:correlation} \langle
S_{n+1}|S_{j}\rangle&=&\frac{1}{2\pi}\int\limits_{0}^{2\pi}d\theta_{0}e^{i(\theta_{n}-\theta_{0})}=\nonumber\\
&=&\frac{1}{2\pi}\int\limits_{0}^{2\pi}d\theta_{0}e^{i\bigg(\theta_{n-1}+\omega_{\nu}(S_{n-1})-K_{0}\sin2\theta_{n-1}+\frac{K_{0}\varepsilon}{2}\sin4\theta_{n-1}-\theta_{0}\bigg)}
\end{eqnarray}
and can be calculated using the above iterative procedure as well as the expression for the
Bessel function
$\exp(iz\sin\varphi)=\sum\limits_{m=-\infty}^{+\infty}J_{m}(z)e^{im\varphi}$.
Taking into account that  $\omega_{\nu}(S_{n})=\omega\approx
const$,  $K_{0}=\beta \omega'(S_{n})\approx const$, from
Eq.~(\ref{eq:correlation}) we deduce
\begin{eqnarray}\label{eq:correlation1}
&&\langle
S_{j+n}|S_{j}\rangle=\frac{1}{2\pi}\int\limits_{0}^{2\pi}d\theta_{0}e^{i(\theta_{n}-\theta_{0})}=\nonumber\\
&&=\frac{1}{2\pi}\int\limits_{0}^{2\pi}d\theta_{0}e^{i(\theta_{n-1}-\theta_{0})}\cdot
e^{-iK_{0}\sin 2\theta_{n-1}}\cdot
e^{\frac{iK_{0}\varepsilon}{2}\sin 4\theta_{n-1}}=\nonumber\\
&&=\frac{1}{2\pi}\int\limits_{0}^{2\pi}d\theta_{0}e^{i(\theta_{n-1}-\theta_{0})}\cdot\sum\limits_{m_{1}=-\infty}^{+\infty}
J_{m_{1}}(K_{0})e^{-2im_{1}\theta_{n-1}}\cdot
\sum\limits_{l_{1}=-\infty}^{+\infty}J_{l_{1}}\bigg(\frac{K_{0}\varepsilon}{2}\bigg)e^{4il_{1}\theta_{n-1}}=\\
&&=e^{in\omega}\sum\limits_{m_{1}=-\infty}^{+\infty}\sum\limits_{m_{2}=-\infty}^{+\infty}\ldots
\sum\limits_{m_{n}=-\infty}^{+\infty}\sum\limits_{l_{1}=-\infty}^{+\infty}\sum\limits_{l_{2}=-\infty}^{+\infty}\ldots
\sum\limits_{l_{n}=-\infty}^{+\infty}(-1)^{l_{1}+l_{2}+\ldots+l_{n}}\cdot
\nonumber
\\&&\cdot e^{-2i\omega m_{1}}e^{-2i\omega(m_{1}+m_{2})}\ldots e^{-2i\omega
( m_{1}+ m_{2}+\ldots+ m_{n-1})}e^{-2i\omega
l_{1}}e^{-2i\omega(l_{1}+l_{2})}\ldots e^{-2i\omega ( l_{1}+
l_{2}+\ldots+ l_{n-1})}\cdot\nonumber\\&& \cdot
J_{m_{1}}[K_{0}]J_{m_{2}}[(1-2m_{1}+4l_{1})K_{0}]\ldots
J_{m_{n}}[(1-2(m_{1}+m_{2}\ldots+m_{n-1})+4(l_{1}+l_{2}\ldots+l_{n-1}))K_{0}]\cdot\nonumber\\
&&\cdot J_{l_{1}}\bigg[\frac{K_{0}\varepsilon}{2}\bigg]\ldots
J_{l_{n}}\bigg[\frac{1-2(m_{1}+m_{2}\ldots+m_{n-1})+4(l_{1}+l_{2}\ldots+l_{n-1}))K_{0}\varepsilon}{2}\bigg]\cdot\nonumber\\
&&\cdot \delta_{1-2(m_{1}+m_{2}+\ldots
+m_{n})+4(l_{1}+l_{2}+\ldots+l_{n});1} \, . \nonumber
\end{eqnarray}
For the details of the derivation of Eq.(24) see \textbf{Appendix
C}.

In the case of a large Lyapunov exponent $J_{m}(K_{0})\sim
(K_{0})^{-1/2}$,   $K_{0}\gg 1$ we infer  from
(\ref{eq:correlation1}) \label{eq:correlation2} \be \langle
S_{j+n}|S_{j}\rangle \sim
\frac{e^{in\omega}}{(K_{0}^{2}\varepsilon/2)^{n/2}}=\exp\bigg(-\frac{n}{\tau_{c}}\bigg)
e^{in\omega} \, , \ee where
$\tau_{c}=2/\ln\bigg(\frac{K_{0}^{2}\varepsilon}{2}\bigg)$ is the
correlation length.  Since $K_{0}\sim \beta$, we have the
following estimation \be\label{eq:tc}
\tau_{c}\sim\frac{2}{\ln\bigg(\frac{\beta^2\varepsilon}{2}\bigg)}
\, . \ee In the isotropic case $\varepsilon=0$, one can perform
the same calculations [insertion of $\varepsilon=0$ into
Eq.~(\ref{eq:tc}) gives a wrong result] and show that
\be\label{eq:tc1} \tau'_{c}\sim\frac{2}{\ln\beta}\,  \ee holds.
Taking into account Eqs.~(\ref{eq:tc}), (\ref{eq:tc1}) and
expressions for the rescaled interaction constants
$\varepsilon\rightarrow
\frac{J-g}{2g},~~\beta\rightarrow\frac{\beta}{g}$ we conclude that
the role of the anisotropy is not trivial. Namely, the strong
anisotropy \be\label{eq:J} J-g>\frac{4g^2}{\beta},
 \ee
 suppresses the spin
correlations because then $\tau'_{c}>\tau_{c}$. On the other hand the
weak anisotropy
\be\label{eq:J1}
J-g<\frac{4g^2}{\beta},
 \ee
 enhances the correlations $\tau'_{c}<\tau_{c}$.

It should be stressed that the reliability of the
 analytical estimates is limited. This is particularly
apparent for the case where the numerical and the analytical
predictions deviate from each others, due to limited range of
applicability of the analytical expressions, derived after rough
approximations.

The role of the anisotropy  $\varepsilon=(J-g)/2g$ can be
clarified numerically as well. In order to better understand the
physical features of the model (\ref{eq:hamiltonian}) we will
study the phase portrait of the system. The results of the
numerical evaluation of the recurrence relations
(\ref{eq:explicit}) are presented in
Figs.~\ref{Fig:2}-\ref{Fig:5}.
\begin{figure}[t]
 \centering
  \includegraphics[width=10cm]{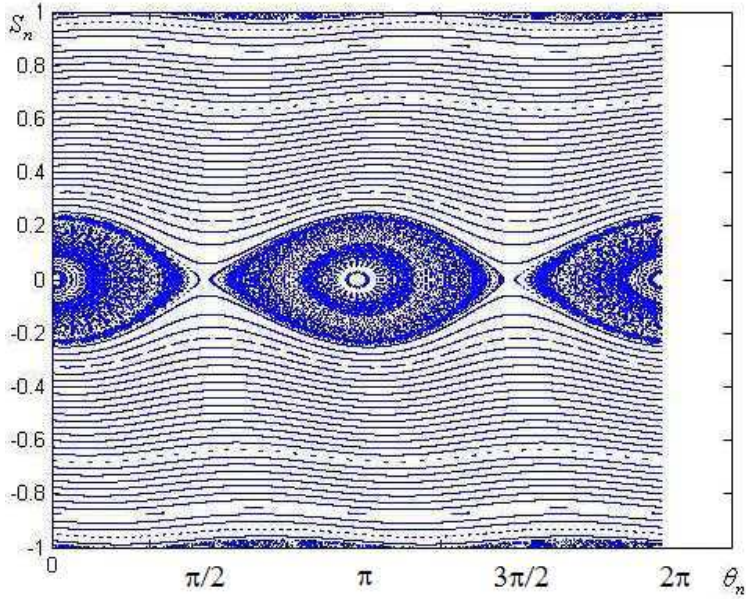}
  \caption{Color online. Results of the numerical calculations of the recurrence relations (\ref{eq:explicit}) on the phase plane
  $(S_{n},\theta_{n})$. The following the parameters are used $\beta=0.05$, $\varepsilon=0.005$.
  About hundred trajectories are generated for the set of different initial conditions $(S_{0},\theta_{0})$.} \label{Fig:2}
\end{figure}
\begin{figure}[t]
 \centering
  \includegraphics[width=10cm]{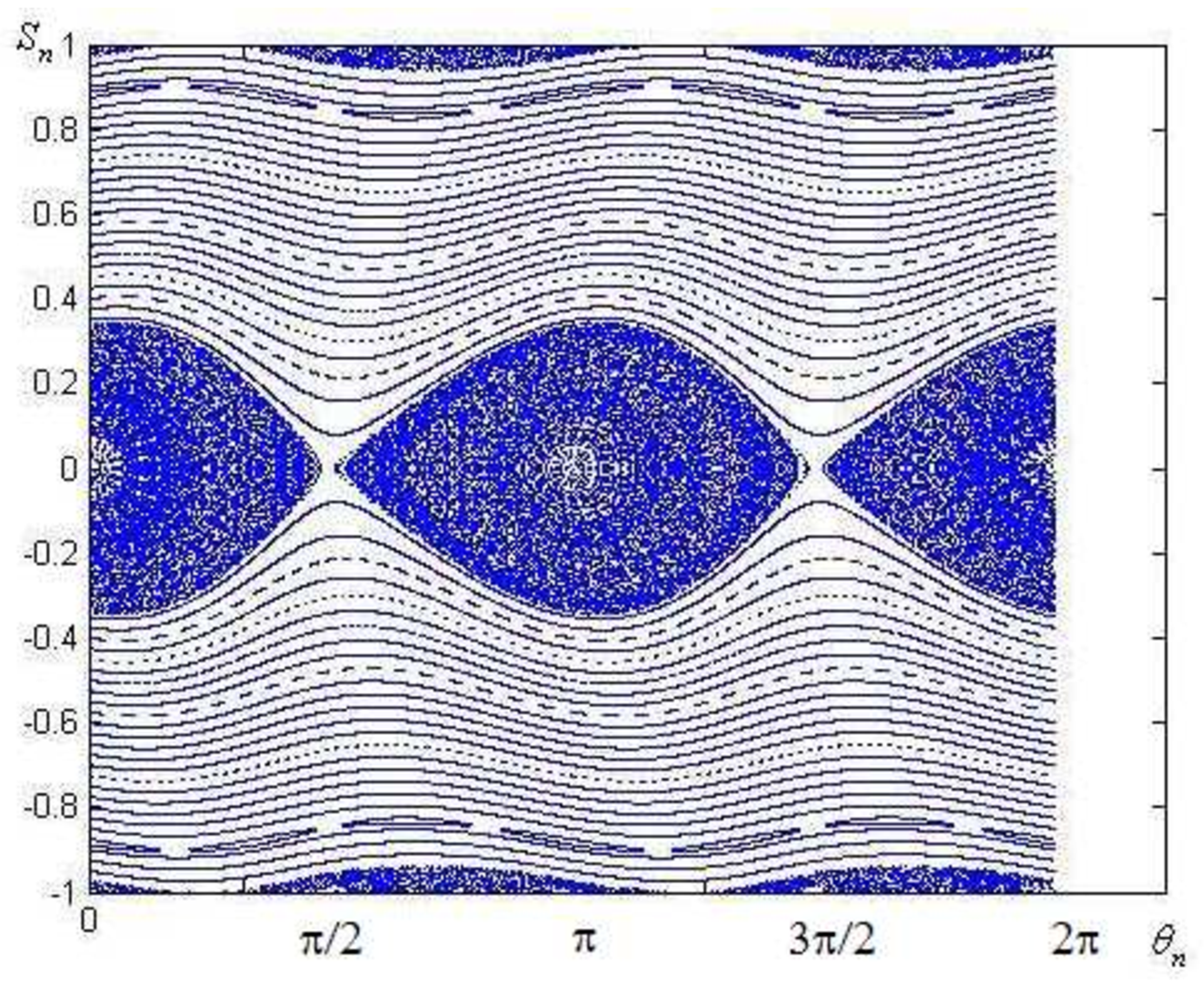}
  \caption{Color online. Results of the numerical calculations of the recurrence relations (\ref{eq:explicit}) on the phase plane
  $(S_{n},\theta_{n})$, for following  the parameters: $\beta=0.05$, $\varepsilon=0$.
  About hundred trajectories are generated for the set of different initial conditions $(S_{0},\theta_{0})$.} \label{Fig:3}
\end{figure}
\begin{figure}[t]
 \centering
  \includegraphics[width=10cm]{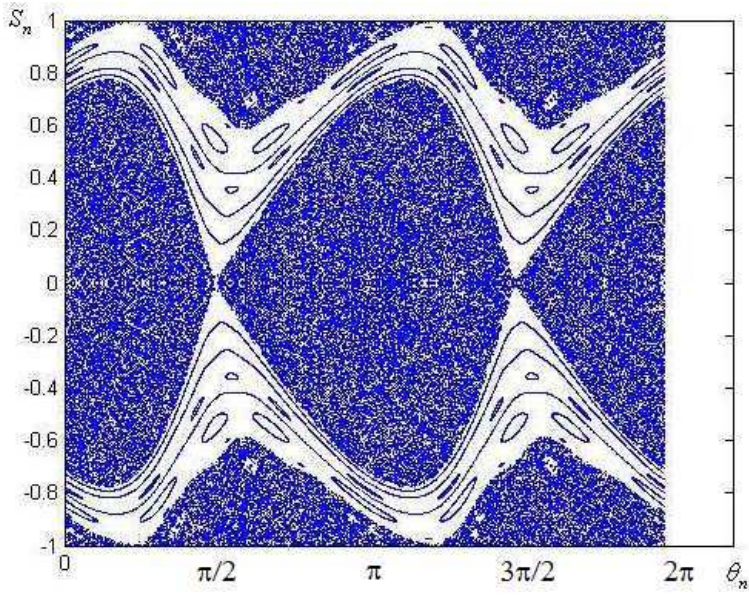}
  \caption{Color online. Results of the numerical calculations of the recurrence relations (\ref{eq:explicit}) on the phase plane
  $(S_{n},\theta_{n})$. We used $\beta=0.3$, $\varepsilon=0.15$.
  About hundred trajectories are generated for the set of different initial conditions $(S_{0},\theta_{0})$.} \label{Fig:4}
\end{figure}
\begin{figure}[t]
 \centering
  \includegraphics[width=10cm]{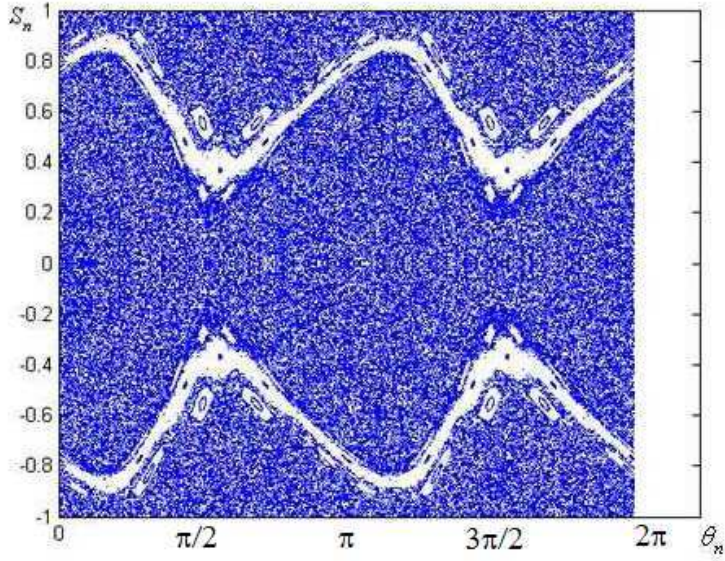}
  \caption{Color online. Results of the numerical calculations of the recurrence relations (\ref{eq:explicit}) on the phase plane
  $(S_{n},\theta_{n})$, for the following  parameters: $\beta=0.3$, $\varepsilon=0$.
  About hundred trajectories are generated for the set of different initial conditions $(S_{0},\theta_{0})$.
  With the increase of the value of parameter $\beta$, the domain of the chaotic motion covers almost the whole  phase space (See Fig. \ref{Fig:4}, Fig.\ref{Fig:5}),
  because the chaos parameter $K$ Eq.(18) is proportional to the constant $\beta$.} \label{Fig:5}
\end{figure}
\begin{figure}[t]
 \centering
  \includegraphics[width=10cm]{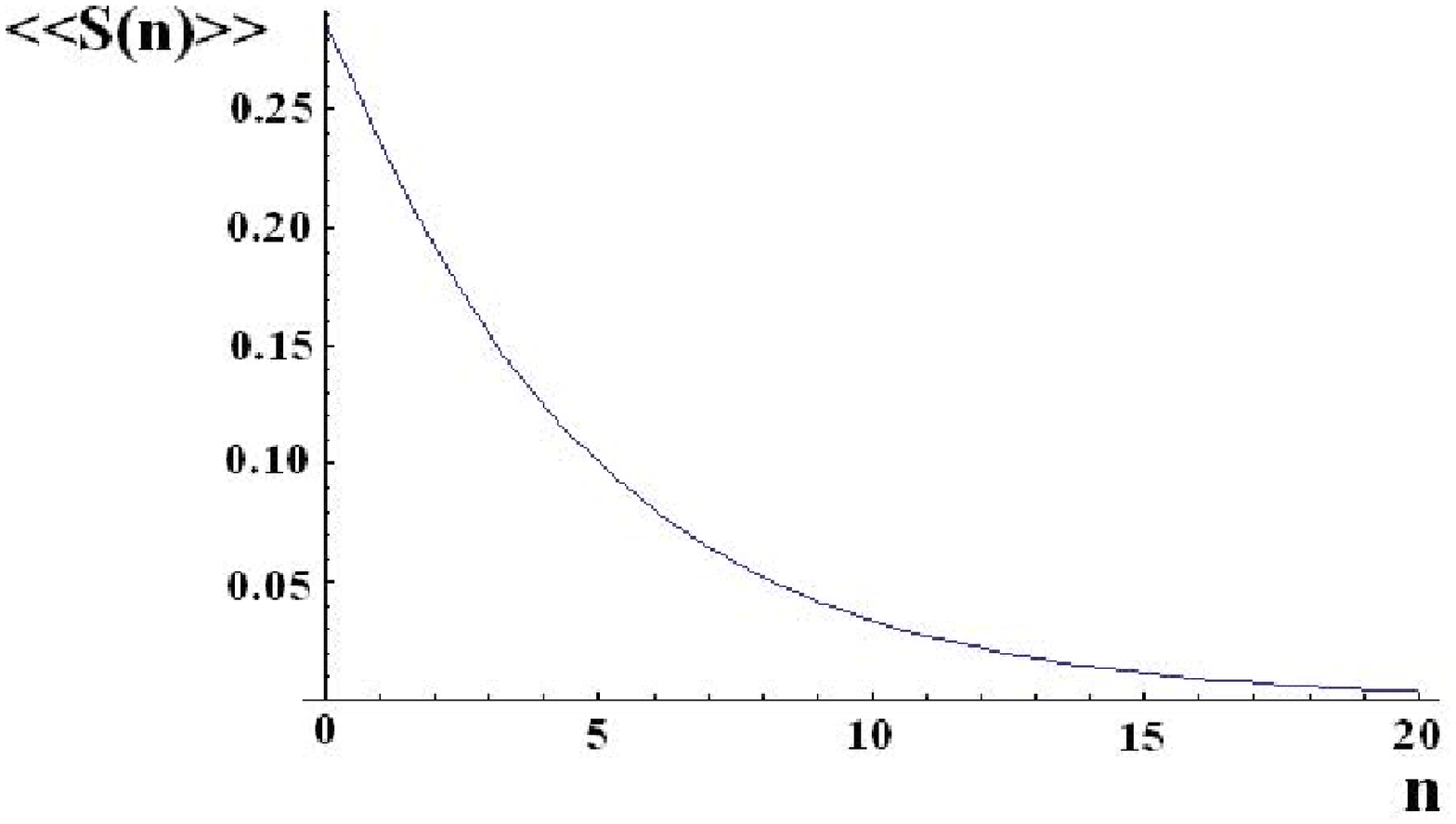}
  \caption{Results of the numerical integration for the statistically averaged random alignment factor
  $\ll\sin^{2}(\theta_{n}-\theta_{n-1})\gg$ in the diffusive approximation (\ref{eq:S2}),
  for $\beta=0.3$ and $\varepsilon=0.15$.} \label{Fig:6}
\end{figure}
As we see from Fig.~\ref{Fig:2} the phase space of the system consists
of two topologically different domains separated from each
other by a separatrix. Most of the phase space belongs to the
domain of the regular motion and open phase trajectories. The domain of
closed phase trajectories mainly corresponds to the irregular
motion and a small island of a regular motion is
observed only in the center of the portrait.
From this formal mathematical statement, one can extract
interesting physical information. Closed trajectories belong
to the oscillatory regime and open trajectories to the rotational one.
Therefore, we expect that two types of motion can be realized for the model (\ref{eq:hamiltonian}). The domain of the
regular spin rotational motion is defined by the relation
$0.2<|S_{n}|<1$.  If then $|S_{n}|<0.2$ the spin oscillation is
chaotic and only for very small amplitude an island of regular
oscillations is observed in the center. If the anisotropy parameter is
zero $\varepsilon=0$ then the island of the regular oscillatory motion
disappears (see Fig.~\ref{Fig:3}). This means that without the small
anisotropy the spin system is less correlated [see  Eqs.~(\ref{eq:J}) and
(\ref{eq:J1})]. Such a geometrical interpretation can be
extrapolated from the pair of the canonical variables $(S_{n},
\theta_{n})$ to the real spin variables $S_{n}^{z}=\cos\theta_{n}$
using the parametrization (\ref{eq:parametrization}) and a simple
relation $S_{n}=\sin(\theta_{n}-\theta_{n-1})$.

\section{Spin diffusion and kinetic approach}

The dynamical picture does not apply in the chaotic regime for
$K>0$ or $K<-4$. An adequate language in this case is the
statistical approach. Instead of the dynamical variables the key
role is played  by the probability distribution function, which is
a solution of the Fokker-Planck equation. Its derivation is rather
straightforward for chaotic dynamical models and is based on the
Kolmogorov, Arnold, Moser (KAM) theory.\cite{Berakdar} Interested
reader can find all technical details of the derivation for the
spin chain model in the recent work \cite{Iomin}. Here we are
using the final result adopted  to the SMM system. The probability
distribution of the spin variable
$S_{n}=\sin(\theta_{n}-\theta_{n-1})$ is described by following
diffusion equation: \be\label{eq:diffusion} \frac{\partial
f}{\partial t}=D\frac{\partial^2 f}{\partial S^2} \, , \ee where
$D(S)=\frac{\beta^2}{4}\bigg(1+\frac{\varepsilon^2}{4}\bigg)$ is
the diffusion coefficient. For the details of derivation of
Eq.(30) see \textbf{Appendix B}. The fundamental solution of this
equation is \be\label{eq:soldiff} f(S,t)=\frac{1}{2\sqrt{\pi D
t}}\exp\bigg(-\frac{S^2}{4Dt}\bigg), \ee and can be found in many
classical textbooks (see e.~g. \cite{Tikhonov}). This solution
(\ref{eq:soldiff}) is defined on the interval $-\infty<S<\infty$
whereas we need one for $-1\leq S\leq1$. In order to find a
solution relevant to our problem we will consider the following
boundary and initial conditions for the diffusion equation
(\ref{eq:diffusion}).
\begin{eqnarray}\label{eq:boundary}
&&f=W_{0}~~~~~\mbox{for}~~~~~t=0,\nonumber\\
&&f=g_{1}(t)~~~~~\mbox{for}~~~~~S=-1,\\
&&f=g_{2}(t)~~~~~\mbox{for}~~~~~S=1,\nonumber
\end{eqnarray}
and will look for the solution in the following form
\be\label{eq:diffsol1} f(S,t)=2\sum\limits_{m=1}^{\infty}\sin(m\pi
S)\exp\big(-Dm^2\pi^2t\big)M_{m}(t), \ee where \be\label{eq:Mt}
M_{m}(t)=\int\limits_{0}^{1}f_{0}(\xi)\sin(n\pi
\xi)d\xi+Dm\pi\int\limits_{0}^{1}\exp\big(Dm^2\pi^2\tau\big)\big[g_{1}(\tau)-(-1)^mg_{2}(\tau)\big]d\tau.
 \ee
 In the simple case $g_{1}(t)=g_{2}(t)=0$  we obtain from Eqs.~(\ref{eq:diffusion})-(\ref{eq:Mt})
 \be\label{eq:simplest}
 f(S,t)=\frac{4W_{0}}{\pi}\sum\limits_{m=0}^{\infty}\frac{1}{(2m+1)}\sin\big((2m+1)\pi S\big)\exp\big(-D(2m+1)\pi^2t\big),
 \ee
 where the coefficient $W_{0}$ can be defined from the normalization
 condition
 $\int\limits_{-1}^{+1}\int\limits_{0}^{+\infty}f(S,t)dSdt=1$,
 $W_{0}=\frac{D\pi^4}{7\varsigma(3)}$. Here
 $\varsigma(s)=\sum\limits_{k=1}^{\infty}k^{-s}$ is the Riemann zeta function
 \cite{Abramowitz}.
     We note the direct correspondence between the fictitious time and the spin index
 $t\rightarrow n T$, $T=1$ Eq.~(\ref{eq:hamsystem}).
For the averages of the discrete random variable $S\equiv
S_{n}=\sin(\theta_{n}-\theta_{n-1})$ we  follow the standard
procedure (see
 \cite{Chotorlishvili}, Eq.(18)-(20)) and utilize the distribution function Eq.~(\ref{eq:simplest}).
 The integration is performed over the interval $-1\leqslant S_{n} \leqslant
 1$. As a result we obtain:
 \begin{eqnarray}
 &&\ll S^{2}_{n}\gg=\ll
 S^{2}\gg=\ll\sin^2(\theta_{n}-\theta_{n-1})\gg=\int\limits_{-1}^{+1}S^2f(S,t)dS=\nonumber\\
 &&=\frac{4W_{0}}{\pi}\sum\limits_{m=0}^{\infty}\frac{1}{(2m+1)}\int\limits_{-1}^{+1}S^2\sin\big((2m+1)\pi|S|\big)dS\cdot
 \exp\big(-D(2m+1)\pi^2t\big).
 \end{eqnarray}
After an integration we get:
 \begin{eqnarray}\label{eq:S2}
 &&\ll S^2\gg=\ll\sin^2(\theta_{n}-\theta_{n-1})\gg=\nonumber\\
 &&=\frac{4W_{0}}{\pi^2}e^{-D\pi^2n}F\bigg(\bigg\{\frac{1}{2},\frac{1}{2},1\bigg\},\bigg\{\frac{3}{2},\frac{3}{2}\bigg\},
 e^{-2D\pi^2n}\bigg)-\frac{W_{0}}{\pi^4}e^{-D\pi^2n}\Phi\bigg(e^{-2D\pi^{2}n},4,\frac{1}{2}\bigg) \, ,
 \end{eqnarray}
 where
 $F(\{a_{1}\ldots a_{p}\};\{b_{1}\ldots b_{q}\};z)=\sum\limits_{k=0}^{\infty}\frac{(a_{1})_{k}\ldots(a_{p})_{k}}{(b_{1})_{k}\ldots(b_{q})_{k}}
 \cdot\frac{z^k}{k!}$ is the generalized hypergeometric function and
 $\Phi(z,s,a)=\sum\limits_{k=0}^{\infty}\frac{z^k}{(k+a)^s}$ is the generalized Riemann zeta function.
 \cite{Abramowitz} From Eq.~(\ref{eq:S2}) we immediately see that the statistically averaged random alignment factor
 $\ll\sin^{2}(\theta_{n}-\theta_{n-1})\gg$ is not uniform along the spin chain (see Fig.~\ref{Fig:6}), but rather decays exponentially with
 $n$. This result is reasonable since the solution for the distribution function (\ref{eq:simplest})
 is obtained via  deterministic initial and boundary conditions (\ref{eq:boundary}). Therefore, a maximum of correlation is
 expected for $n=0$. Since $t=nT$, $t=0$,
 $n=0$ corresponds to the boundary where the distribution function is
 defined precisely. Far away from the boundary that means for
 $n>>1$  randomness occurs and the correlation decays.

\section{Conclusions}
In this paper we considered an anisotropic nonlinear spin chain,
which serves as a model for a  chain of coupled
nanomagnets. We have shown, that there is a direct map between an
infinite-dimensional spin chain model and an equivalent effective
$1D$ classical Hamiltonian  with a discrete fictitious
time-dependent perturbation. We have established a direct
connection between the dynamical characteristics of the classical
system and the spin correlation time of the original quantum
chain. The decay rate for the spin correlation functions turns out to
 depend logarithmically on the maximal Lyapunov exponent. In
addition, for an anisotropic couplings we found an interesting
counterintuitive feature: the small anisotropy leads to the
formation of small islands of the regular motion in a chaotic sea
of the system's phase space. As a result, the spin correlations
become stronger within the islands of regular motion. We argue
that these results obtained within the classical approximation are
interesting in other regimes. If the Lyapunov regime is reached
for a quantum system, which takes place for the Lyapunov exponent
$\lambda< J^2/\Delta$, where $\Delta$ is the mean level spacing
and $J$ is the exchange interaction constant between spins, then
the decay rate for the teleportation fidelity in a device based on
 such spin chains is directly related to $\lambda$.

\textbf{Acknowledgments:} We thank Boris Fine for useful
discussions. The financial support by the Deutsche
Forschungsgemeinschaft (DFG) through SFB 762, contract BE 2161/5-1, the Grant No.
KO-2235/3, and STCU Grant No. 5053 is gratefully acknowledged.

\appendix
\section{Derivation of the recurrence relations}
Let us consider the equilibrium state for the model (\ref{eq:ham}):
\begin{eqnarray}&& \frac{\partial H}{\partial \varphi
_{n}}=\bigg(\cos(\theta_{n}-\theta_{n+1})-\cos(\theta_{n}+\theta_{n+1})\bigg)\sin(\varphi_{n+1}-\varphi_{n})-\nonumber\\
&&-\bigg(\cos(\theta_{n-1}-\theta_{n})-\cos(\theta_{n-1}+\theta_{n})\bigg)\sin(\varphi_{n+1}-\varphi_{n}),\label{eq:dhdfi}
\end{eqnarray}
\begin{eqnarray}
&&\frac{\partial H}{\partial
\theta_{n}}=-\frac{J}{2}\bigg(\sin(\theta_{n}+\theta_{n+1})+\sin(\theta_{n-1}+\theta_{n})+\sin(\theta_{n}-\theta_{n+1})-\sin(\theta_{n-1}+\theta_{n})-\nonumber\\
&&-\frac{g}{2}\bigg(\sin(\theta_{n}-\theta_{n+1})-\sin(\theta_{n-1}-\theta_{n})-\sin(\theta_{n}+\theta_{n+1})-\sin(\theta_{n-1}+\theta_{n})\bigg)\cos(\varphi_{n+1}-\varphi_{n})+\nonumber\\
&&+\beta\sin 2\theta_{n}=0.\label{eq:dhdtheta}
\end{eqnarray}
After introduction of the notation $S_{n}=\sin(\theta_{n}-\theta_{n-1})$,
from (\ref{eq:dhdfi}), (\ref{eq:dhdtheta}) we find:
\begin{eqnarray}\label{eq:A3}
&&\Bigg(\frac{J}{2}+\frac{g}{2}\Bigg)S_{n+1}-\Bigg(\frac{J}{2}+\frac{g}{2}\Bigg)S_{n}-\Bigg(\frac{J}{2}-\frac{g}{2}\Bigg)\sin(\theta_{n}+\theta_{n+1})-\nonumber\\
&&-\Bigg(\frac{J}{2}-\frac{g}{2}\Bigg)\sin(\theta_{n-1}+\theta_{n})+\beta\sin(2\theta_{n})=0,\\
&&\varphi_{n+1}=\varphi_{n}+\pi m,~~~m=0,1\nonumber \, .
\end{eqnarray}
Let the asymmetry parameter be defined by $\varepsilon=|J-g|$,
$\varepsilon<J,g$. Next we perform a rescaling of the interaction
constants $\frac{\varepsilon}{2g}\rightarrow\varepsilon$,
$\beta\rightarrow \mp \frac{\beta}{g}$. From (\ref{eq:A3}) we
deduce:
\begin{eqnarray}\label{eq:A4}
(S_{n+1}-S_{n})-\varepsilon
\sin(\theta_{n}+\theta_{n+1})\varepsilon
-\sin(\theta_{n-1}+\theta_{n})+\beta\sin(2\theta_{n})=0,\\
\theta_{n+1}=\theta_{n}+\pi
\nu+(-1)^{\nu}\arcsin[S_{n+1}],~~~\nu=0,1\nonumber \, .
\end{eqnarray}
Depending on the sign of the rescaled barrier height
$\beta\rightarrow\mp\frac{\beta}{g}$, the value of the index $m=0,1$
defines the energy minimum condition. For convenience we will use
positively defined  $\beta>0$ and consequently $m=0$.   In the
simplest case  $\nu=0$, so that from (\ref{eq:A4}) we obtain:
\begin{eqnarray}\label{eq:A5}
&&(S_{n+1}-S_{n})\nonumber\\
&&-\varepsilon
\Bigg(\sin[2\theta_{n}]\sqrt{1-S_{n+1}^2}+\cos(2\theta_{n})S_{n+1}+\sin(2\theta_{n})\sqrt{1-S_{n}^2}-\cos(2\theta_{n})S_{n+1}\Bigg)+\\
&&+\beta\sin(2\theta_{n})=0 \, .\nonumber
\end{eqnarray}
Retaining only the first order terms with respect to the small parameter
$\varepsilon=\frac{|J-g|}{2g}$, from (\ref{eq:A5}) we
obtain the following recurrence relations (\ref{eq:explicit}):
\begin{eqnarray}\label{eq:A6}
&&S_{n+1}=S_{n}-\beta\bigg(\sin2\theta_{n}-\frac{\varepsilon}{2}\sin
4\theta_{n}\bigg), \nonumber\\
&&\theta_{n+1}=\theta_{n}+\omega_{\nu}(S_{n+1}),\\
&&\omega_{\nu}(S_{n})=\pi \nu +(-1)^{\nu}\arcsin S_{n} \, . \nonumber
\end{eqnarray}

\section{Derivation of the kinetic equation}
The starting point for the derivation of the kinetic equation is the
equivalent effective Hamiltonian (\ref{eq:hamsystem}):
\begin{eqnarray}\label{eq:B1}
&&\frac{dS}{dt}=-\frac{\partial H}{\partial \theta}=-\beta
V'(\theta)T
\sum\limits_{n=-\infty}^{+\infty}\delta(t-n T),\nonumber\\
&& \frac{d \theta}{d t}=\frac{\partial H}{\partial S}=\omega (S),
\\
&& \omega_{\nu}(S)=\pi \nu+(-1)^{\nu}\arcsin s \, . \nonumber
\end{eqnarray}
Here the variable $S$ plays the role of the adiabatic (slowly varying) action variable, while the angular variable $\theta$  is the fast variable.
Due to the presence of the two different time scales in the
system: \be H=H_{0}(S)+\varepsilon V(S,\theta, t)\label{eq:B2}
 \ee
 for the derivation of the  kinetic equation we will follow to the standard procedure.
 \cite{Toklikishvili}  The distribution function of the random
 variable $f(S,t)$ obeys the  Liouvillian equation of motion:
 \begin{eqnarray}\label{eq:B3}
 && i\frac{\partial f_{0}}{\partial t}=(\hat{L}_{0}+\varepsilon
 \hat{L}_{1})f_{0}, \nonumber\\
 &&\hat{L}_{0}=i\omega(S)\frac{\partial}{\partial \theta}, \\
&&\hat{L}_{1}=-i\Bigg(\frac{\partial V}{\partial
S}\frac{\partial}{\partial\theta}-\frac{\partial V}{\partial
S}\frac{\partial}{\partial\theta}\Bigg).\nonumber
 \end{eqnarray}
The formal solution of the Liouville equation with the accuracy of second order terms in the small parameter  $\varepsilon$ reads:
  \begin{eqnarray}\label{eq:B4}
  &&f_{0}(S,t)=f(S,0)-i\varepsilon\sum\limits_{m}\int\limits_{0}^{t}dt_{1}\exp[im\int\limits_{0}^{t_{1}}\omega(t')dt']\langle
  n|\hat{L}_{1}|m\rangle f_{0}(S,0)+\nonumber\\
  &&+(-i\varepsilon)^{2}\sum_{m}\int\limits_{0}^{t}dt_{1}\int\limits_{0}^{t_{1}}dt_{2}\exp[-im\int\limits_{t_{1}}^{t_{2}}\omega(t')dt']\langle0|\hat{L}_{1}(t_{1})|m\rangle\langle
  m|\hat{L}_{1}(t_{2})|0\rangle f_{0}(S,0).
  \end{eqnarray}
  Here $\langle
  n|\hat{L}_{1}|m\rangle\frac{1}{2\pi}\int\limits_{0}^{2\pi}d\theta e^{-in\theta}\hat{L}_{1}e^{im\theta}$
  is the matrix element of the Liouville operator.  After averaging over the initial
  phases $f(I,t)=\ll f_{0}(I,t)\gg$ and applying the random phase approximation with respect to the fast chaotic variable
  $\Psi(t_{2},t_{1})=\int\limits_{t_{1}}^{t_{2}}\omega(t')dt'=\theta(t_{1})-\theta(t_{2})$
  \be\label{eq:b5}
  \ll\exp i m
  \Psi(t_{2},t_{1})\gg\approx\exp(-(t_{1}-t_{2})/\tau_{c})\exp(-im\omega(t_{1}-t_{2})) \, .
  \ee
  From (\ref{eq:B4}) we obtain:
  \be\label{eq:B6}
  \frac{\partial f}{\partial
  t}=-2\varepsilon^2\sum\limits_{m>0}\sum\limits_{p>0}\frac{(1/\tau_{c})\langle0|\hat{L}_{1p}|m\rangle\langle m|\hat{L}_{1-p}|0\rangle f}{(1/\tau_{c})^2+(m\omega-p\Omega)^2} \, ,
 \ee
 where
 \begin{eqnarray}\label{eq:B7}
 &&\langle0|\hat{L}_{1p}|m\rangle\langle
 m|\hat{L}_{1-p}|0\rangle=\nonumber\\
 &&=\bigg(\frac{\Omega}{2\pi}\bigg)^2\frac{1}{(2\pi)^2}\int\limits_{-T/2}^{T/2}dt_{1}\int\limits_{-T/2}^{T/2}dt_{2}\int\limits_{0}^{2\pi}d\theta'\int\limits
 d\theta"
 \hat{L}_{1}(t_{1})e^{im(\theta'-\theta")}\hat{L}_{1}(t_{2})e^{-ip\Omega(t_{1}-t_{2})}
 \end{eqnarray}
 and the following notation is used $\tau_{c}=2T/\ln K$,
 $T=\frac{2\pi}{\Omega}$.  After calculating the integrals in (\ref{eq:B7}), in
 the limit $\frac{1}{\tau_{c\Omega}}\rightarrow 0$, $T=1$ form (\ref{eq:B6}) we simply recover the diffusion equation (\ref{eq:diffusion}):
 \begin{eqnarray}\label{eq:B9}
&& \frac{\partial f(S,t)}{\partial t}=\frac{\partial}{\partial
S}D(S)\frac{\partial f(S,t)}{\partial
 S}, \\
&& D(S)=\frac{\beta^2}{4}\Bigg(1+\frac{\varepsilon^2}{4}\Bigg) \, .
 \end{eqnarray}
 More details of derivations can be found in  Ref.~\onlinecite{Toklikishvili}

 \section{Correlation functions}
 For the evaluation of the multiple series in Eq.~(\ref{eq:correlation1}) one should sum
 up the contributions from the main non-oscillatory terms. Due to the delta
 function
 $\delta_{1-2(m_{1}+m_{2}+\ldots+m_{n})+4(l_{1}+l_{2}+\ldots+l_{n}),1}$ in Eq.~(\ref{eq:correlation1}), and the fast exponential factors
 $e^{i4\omega(l_{1}+l_{2}+\ldots+l_{n})}$,
 $e^{-2i\omega(m_{1}+m_{2}+\ldots+m_{n})}$, the relevant terms in Eq.~(\ref{eq:correlation1})  are those with
 \begin{eqnarray}\label{eq:C1}
 m_{1}+m_{2}+\ldots+m_{n}=0,\nonumber \\
 l_{1}+l_{2}+\ldots +l_{n}=0.
 \end{eqnarray}
Using the asymptotic expressions for Bessel functions: \be
J_{m}(K_{0})\sim K_{0}^{-1/2},~~\mbox{for}~~K_{0}\gg 1,
 \ee
 and condition (\ref{eq:C1}), one can easily obtain (26) from (\ref{eq:correlation1}).
\end{widetext}

\end{document}